\documentclass[compsoc, conference, a4paper, 10pt, times]{IEEEtran}

\usepackage{packages}
\DeclareAcronym{hsts}{
    short = HSTS,
    long = HTTP Strict Transport Security
}

\DeclareAcronym{ct}{
    short = CT,
    long = Certificate Transparency
}

\DeclareAcronym{ech}{
    short = ECH,
    long = Encrypted Client Hello
}

\DeclareAcronym{cctld}{
    short = ccTLD,
    long = country-code TLD
}

\DeclareAcronym{ip}{
    short = IP,
    long = Internet Protocol
}

\DeclareAcronym{cdf}{
    short = CDF,
    long = Cummulative Distribution Function
}

\DeclareAcronym{ppc}{
    short = PPC,
    long = Pay-Per-Click
}

\DeclareAcronym{ppr}{
    short = PPR,
    long = Pay-Per-Redirect
}

\DeclareAcronym{mss}{
    short = MSS,
    long = Maximum Segment Size
}

\DeclareAcronym{iid}{
    short = IID,
    long = IPv6 Interface ID
}

\DeclareAcronym{oui}{
    short = OUI,
    long = Organizationally Unique Identifier
}

\DeclareAcronym{tbt}{
    short = TBT,
    long = Too Big Trick
}

\DeclareAcronym{isp}{
    short = ISP,
    long = Internet Service Provider
}

\DeclareAcronym{cpe}{
    short = CPE,
    long = Customer-premises equipment
}

\DeclareAcronym{gfw}{
    short = GFW,
    long = Great Firewall of China
}

\DeclareAcronym{rir}{
    short = RIR,
    long = Regional Internet Registry
}

\DeclareAcronym{tld}{
    short = TLD,
    long = Top Level Domain
}

\DeclareAcronym{pop}{
    short = POP,
    long = Point of Presence
}

\DeclareAcronym{pmtu}{
    short = PMTU,
    long = Path Maximum Transmission Unit
}

\DeclareAcronym{cdn}{
    short = CDN,
    long = Content Delivery Network
}

\DeclareAcronym{as}{
    short = AS,
    long = Autonomous System,
    short-plural-form = ASes
}

\DeclareAcronym{alpn}{
    short = ALPN,
    long = Application-Layer Protocol Negotiation
}

\DeclareAcronym{czds}{
    short = CZDS,
    long = Centralized Zone Data Service
}

\DeclareAcronym{sni}{
    short = SNI,
    long = Server Name Indication
}

\DeclareAcronym{ietf}{
    short = IETF,
    long = Internet Engineering Task Force
}

\DeclareAcronym{dns}{
    short = DNS,
    long = Domain Name System
}

\DeclareAcronym{dnsrr}{
    short = DNS RR,
    long = \acl{dns} Resource Record
}

\DeclareAcronym{tcp}{
    short = TCP,
    long                = {TCP}, %
    first-style         = short,
}

\DeclareAcronym{tls}{
    short = TLS,
    long = Transport Layer Security
}

\DeclareAcronym{http}{
    short = HTTP,
    long = Hypertext Transfer Protocol
}

\DeclareAcronym{https}{
    short = HTTPS,
    long = Hypertext Transfer Protocol Secure
}

\DeclareAcronym{altsvc}{
    short = ALT-SVC,
    long = Alternative Service
}

\crefname{lstlisting}{listing}{listings}
\Crefname{lstlisting}{Listing}{Listings}

\usepackage{xurl}

\makeatletter
\let\old@ps@headings\ps@headings
\let\old@ps@IEEEtitlepagestyle\ps@IEEEtitlepagestyle
\def\confheader#1{%
    \def\ps@IEEEtitlepagestyle{%
        \old@ps@IEEEtitlepagestyle%
        \def\@oddhead{\strut\hfill#1\hfill\strut}%
        \def\@evenhead{\strut\hfill#1\hfill\strut}%
    }%
    \ps@headings%
}
\makeatother

\confheader{%
    \fbox{
        \begin{minipage}{2\linewidth}
            \footnotesize
        If you cite this paper, please use the WTMC reference: J. Zirngibl, P. Sattler, G. Carle. 2023. A First Look at SVCB and HTTPS DNS Resource Records in the Wild. \textit{International Workshop on Traffic Measurements for Cybersecurity 2023}. WTMC.
        \end{minipage}
}
}

\begin{document}

\title{A First Look at SVCB and HTTPS DNS Resource Records in the Wild}

\author{
\IEEEauthorblockN{Johannes Zirngibl, Patrick Sattler, Georg Carle}
\IEEEauthorblockA{
	\textit{Technical University of Munich, Germany}\\
		\{zirngibl, sattler, carle\}@net.in.tum.de\\
	}
}

\maketitle

\begin{abstract}
The Internet Engineering Task Force is standardizing new DNS resource records, namely \svcbrr and \httpsrr.
Both records inform clients about endpoint and service properties such as supported application layer protocols, IP address hints or \ac{ech} information.
Therefore, they allow clients to reduce required DNS queries and potential retries during connection establishment and thus help to improve the quality of experience and privacy of the client.
The latter is achieved by reducing visible meta-data, which is further improved with encrypted DNS and \ac{ech}.

The standardization is in its final stages and companies announced support, \eg Cloudflare and Apple.
Therefore, we provide the first large-scale overview of actual record deployment by analyzing more than \sm{400} domains.
We find \sk{3.96} \svcbrr and \sm{10.5} \httpsrr records.
As of March 2023, Cloudflare hosts and serves most domains, and most records only contain Application-Layer Protocol Negotiation (ALPN) and IP address hints.
Besides Cloudflare, we see adoption by a variety of authoritative name servers and hosting providers indicating increased adoption in the near future.
Lastly, we can verify the correctness of records for more than \sperc{93} of domains based on three application layer scans.
\end{abstract}

\section{Introduction}
\label{sec:introduction}

With the ongoing development of the Internet, available protocols and versions, a general requirement is getting more important, namely \textit{information about supported application layer protocols, versions and properties by individual endpoints}.
The latter information can be exchanged during a handshake or first communication (\eg \ac{altsvc} Headers in \ac{http}).
However, missing knowledge increases the handshake duration and information from existing solutions can only be used in subsequent connections.
Each connection attempt and the potential use of insecure protocols reveals further meta-data related to a client and its desired connection, thus impacting its privacy and security.

To circumvent this problem, the \ac{ietf} works on a new general \ac{dnsrr} named \svcbrr ("SerViCe Binding") that provides service bindings for a domain~\cite{draft-ietf-dnsop-svcb-https-12}.
This record accomplishes two major goals, directing a client \first to another alias or \second to an endpoint including service information.
As a first subtype, the \httpsrr \ac{dnsrr} is specified with a focus on \ac{https} endpoints.
The records allow a client to receive all required information, namely supported protocols, used ports and IP addresses, using a \textit{single}, recursive DNS query.
Provided information can be used to directly establish a secure communication channel using a protocol both endpoints support.
Information about available application protocols and their explicit version can also reduce the risk of on-path or downgrade attacks, \eg make \ac{hsts} obsolete.
Furthermore, the new \httpsrr record is supposed to be extended to provide \ac{ech} information to the client in the future. 
Once specified and deployed, \ac{ech} \cite{draft-ietf-tls-esni-16} further reduces the visibility of connection-related meta-data, \eg the \ac{sni}.

Quick and widespread deployment of these new records can drastically improve the privacy of clients on the Internet.
Different operators including Cloudflare \cite{cloudflare_announcement} and Akamai \cite{akamai_announcement} but also client software, \eg Apple iOS \cite{apple_announcement} and Google Chromium \cite{chromium_announcement} have already announced support for the new records.

\vspace{0.2em}
\noindent
Therefore, we set out to evaluate actual deployments and availability of the new records based on a large-scale measurement.
Our contributions in this paper are:

\first We evaluate the support of new records for more than \sm{400} domains.
We show that the deployment is mostly driven by Cloudflare. However, other operators show initial deployment as well.

\second We evaluate the properties of received records and their implication for a client and established connections.
We show that most domains have records with service information, mainly \ac{alpn} values and \textit{ipv4-} and \textit{ipv6hints}.
Further parameters are rarely visible.

\third We verify the correctness of received information with application layer scans.
We were able to connect to \sperc{96} of targets extracted from \httpsrr records.
\section{Background}
\label{sec:background}

\begin{table*}
	\caption{Example \svcbrr and \httpsrr \acp{dnsrr}}
	\label{tab:httpsrr}
	\centering
	\begin{tabular}{lllllll}
		\toprule
		Domain &		TTL&	CLASS &	TYPE	 & Priority	 &Target Name &	SvcParams	\\
		\midrule
		coffebike.no. &	3600 &	IN &	SVCB &	0 &	barmobile.no. &\\
		cloudflare.com.	 &30 &	IN &	HTTPS &	1 &	. 	 &	alpn="h3,h2" ipv4hint=104.16.132.229,104.16.133.229\\
		\bottomrule
	\end{tabular}
\end{table*}

The \svcbrr \ac{dnsrr} 
represents a more general record to be used with different service types, while the \httpsrr  \ac{dnsrr} is specifically designed to be used with HTTPS.
These \acp{dnsrr} allow clients to select the correct service properties directly.
To indicate the desired service, domains for \svcbrr records should be prefixed with Attrleaf labels~\cite{rfc8552} (\eg \textit{\textunderscore{}dns}).
Using \httpsrr records implies \ac{http} as service.
\Cref{tab:httpsrr} shows two example records.
\ac{ietf} designs both records to be flexible and expandable.
The first \svcbrr record is in alias mode, indicated by the priority of \textit{0}, and redirects the domain to another target name.
In comparison to canonical name (CNAME) records, this is also possible at the apex of a zone~\cite{draft-ietf-dnsop-svcb-https-12}.

The second \httpsrr record is in service mode and provides further information about the endpoint.
In service mode, a target name can be set to indicate another name.
The target name is ”.” if the actual domain should be used.
Additional record data is organized as key-value data, so-called \textit{SvcParams}.
Each parameter has to have a specified format to allow interoperability.
As of March 2023, the draft specifies six different parameter keys and their value format.
By default, an \httpsrr record indicates \ac{http}/1.1 support.
The \textit{alpn} parameter can indicate additional protocols.
If an endpoint does not support \ac{http}/1.1 but other \acp{alpn} the \textit{no-default-alpn} parameter has to be added.
The \textit{port} parameter allows indicating alternative ports, while \textit{ipv4-} and \textit{ipv6hint} allow informing about IP addresses.
Finally, the \textit{mandatory} parameter can be used to indicate a set of parameters that must be used for the service to function correctly.

The initially drafted but now reserved \textit{ech} parameter relies on a different draft \cite{draft-ietf-tls-esni-16}.
However, it lacks deployment (see \Cref{sec:analysis}) and its final publication is delayed.
Therefore, after a discussion~\cite{logjam}, the parameter and references were removed from the \svcbrr and \httpsrr draft \cite{draft-ietf-dnsop-svcb-https-12} to allow an RFC publication.
We evaluate the presence of this parameter in \Cref{sec:analysis}.

For \svcbrr records prefixed with \textit{\textunderscore{}dns}, the respective draft additionally adds the \textit{dohpath} parameter that allows to specify a Uniform Resource Identifier (URI) template for \ac{dns} over \ac{https} \cite{draft-ietf-add-svcb-dns}.

\section{Data Collection}
\label{sec:data}
This work relies on active measurements to collect \ac{dns} data and verify the usefulness of collected records using \ac{http} scans.
This section explains all scans conducted between February 22nd and March 9th, 2023, and covers ethical considerations.

\vspace{0.2em}
\noindent
{\bf DNS Scans}:
We used MassDNS\footnote{\url{https://github.com/blechschmidt/massdns}} with a local Unbound resolver to resolve more than \sm{400} domains to their \svcbrr and \httpsrr, but also A and NS records.
We further resolved the name server domains from the latter NS records to their respective A records.
This allows us to analyze who serves the new record and which operators are involved.
We combined domains from the following sources as input for our measurement:

\noindent
\first Names on the Majestic~\cite{majestic}, Alexa\footnote{We use the last published list before deprecation from February 1st, 2023. \url{https://toplists.net.in.tum.de/archive/alexa/}}~\cite{alexa}, and Umbrella~\cite{umbrella} \topm lists;

\noindent
\second More than \sk{1} available zone files from the \acl{czds}, \eg \textit{.com}, \textit{.net} and \textit{.org};

\noindent
\third A static collection of \sm{98} domains from \num{52} \aclp{cctld} (partial zones, \eg  \sm{13} \textit{.de} domains);

\noindent
\fourth \textit{www.} domains extracted from \acl{ct} logs between August 2022 and January 2023. 

We additionally prefixed domains with the Attrleaf label \textit{\textunderscore{}dns}~\cite{rfc8552}.
As of March 2023, it was the only available label based on an IETF draft \cite{draft-ietf-add-svcb-dns}.
We exclude {www.} domains for this measurement but included domains from NS record names.

\vspace{0.2em}
\noindent
{\bf Protocol Scans}:
We used the \qscanner introduced by Zirngibl \etal~\cite{zirngibl2021over9000} and the Goscanner \cite{goscanner} to test whether received \ac{alpn} information is valid for the given domain.
The \qscanner supports \quic handshakes and \ac{http}/3 requests while the Goscanner supports \ac{tls}/\ac{tcp} handshakes and \ac{http}/1.1 and \ac{http}/2 requests.

For each domain with an \httpsrr record in service mode, we extracted the supported \acp{alpn}, port and IP addresses from the \textit{ipv4hint} in the records.
If no \textit{ipv4hint} is available, we rely on each domain's additionally requested A records.
We use these tuples of domain, IP address, port, and \ac{alpn} to seed our protocol scans.

\vspace{0.2em}
\noindent
{\bf Ethics}:
During all our scans, we strictly followed a set of ethical measures, \ie informed consent~\cite{menloreport} and community best practices~\cite{PA16}. 
Our scans are conducted with a limited rate and use a request-based  blocklist.
Furthermore, our measurement vantage point is clearly identified based on reverse DNS, WHOIS information, and a hosted website. 
We did not receive any inquiries related to our scans during this work.

\section{Analysis}
\label{sec:analysis}

\begin{table*}
	\centering
	\footnotesize
	\caption{Number of domains with each property and parameter in their \svcbrr and \httpsrr DNS resource records.}
	\label{tab:general_results}
	\begin{tabular}{lrrrrrrrrrrr}
		\toprule
		&& \multicolumn{2}{c}{Mode} & \multicolumn{8}{c}{Keys} \\
		\cmidrule(lr){3-4} \cmidrule(lr){5-12}
		Record & Total & Alias & Service &  Mandatory & ALPN & No Default & Port &  ECH &  IPv4 Hint & IPv6 Hint & DoH Path \\
		\midrule
		SVCB & \sk{3.96} & \sk{3.9} & 62 & 0 & 53 & 0 & 2 & 0 & 25 & 15 & - \\
		HTTPS & \sm{10.56} & \sk{2.6} & \sm{10.55} & 0 & \sm{10.55} & 0 & 13 & 20 & \sm{10.55} & \sm{10.23} & - \\
		SVCB + \textit{\textunderscore{}dns} & 27 & 0 & 27 & 0 & 26 & 0 & 12 & 0 & 1 &  1&1 \\ 
		\bottomrule
	\end{tabular}
\end{table*}

We analyze the current deployment of \svcbrr and \httpsrr records based on our measurements described in \Cref{sec:data}.
Resolving more than \sm{400} domains, we received \svcbrr records for \sk{3.96} domains but \httpsrr records for \sm{10.56} domains.
\svcbrr should be available for domains with Attrleaf labels \cite{rfc8552}.
Therefore, we additionally resolved domains prefixed with the first specified label (\textit{\textunderscore{}dns}) but only received records for 27 domains.

\subsection{General Record Analysis}
\Cref{tab:general_results} shows which modes (alias vs service) are used and which keys are commonly present in available records.
Regarding \svcbrr records, \sk{3.9} (\sperc{98.4}) domains use the record for alias mode, aliasing the service to a different domain.
Only 62 domains use the service mode and mostly advertise \ac{alpn} values or IPv4 and IPv6 addresses as hints.
27 domains prefixed with \textit{\textunderscore{}dns} result in \svcbrr records.
All records are in service mode advertising different \ac{alpn} values (4$\times$ \textit{h2} for \ac{dns} over \ac{https} and 26$\times$ \textit{dot} for \ac{dns} over \ac{tls}).
The DoH path advertised by a single domain is \texttt{/dns-query{?dns}}.
The \svcbrr record in both scenarios is only deployed by few domains and we focus on \httpsrr records for the remainder of this paper.

\begin{table}
	\centering
	\footnotesize
	\caption{Top 5 advertised ALPN sets/values in \httpsrr DNS resource records. Note that \httpsrr records imply the support of \ac{http}/1.1 by default \cite{draft-ietf-dnsop-svcb-https-12}.}
	\label{tab:alpn_top5}
	\begin{tabular}{lr|lr}
		\toprule
		ALPN sets & Domains & ALPN values & Domains\\
		\midrule
		h3, h3-29, h2 &  \sm{9.72} & h2 & \sm{10.55}\\
		h2 &   \sm{0.83} & h3 & \sm{9.72} \\
		& \sk{3.23} & h3-29 &  \sm{9.72} \\
		h3, h3-29 &      866 & http/1.1 & 15 \\
		h2, h3 &      242 & h2c & 10\\
		\bottomrule
	\end{tabular}
\end{table}

Regarding \httpsrr records, only \sk{2.6} (\sperc{0.02}) domains use the alias mode, while a majority advertises endpoint information using the service mode.
Similarly, most domains advertise \ac{alpn} values and IPv4 and IPv6 addresses as hints.
The \httpsrr record implies support of \ac{http}/1.1 by default if the \textit{no-default-alpn} parameter is not present.
In our results, no domain  with an \httpsrr record in service mode has the flag set.
\Cref{tab:alpn_top5} shows the Top-5 advertised \ac{alpn} parameters.
A majority of domains advertise \ac{http} version 2 but also 3 indicating \quic support, while \sk{834.4} only advertise \ac{http}/2.
\sk{3.2} domains do not advertise additional \ac{alpn} values but only rely on the default.
A client can still use record information and only establish a connection if it supports \ac{http}/1.1.

\begin{figure}
	\includegraphics{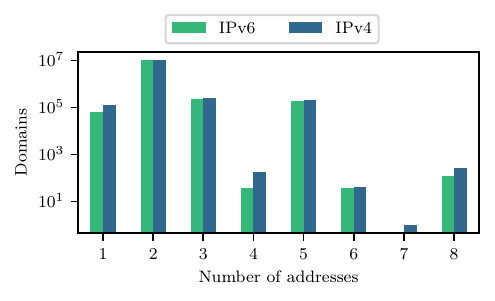}
	\caption{Addresses in \textit{ipv4-} and \textit{ipv6hints}. Note the logarithmic y-axis.}
	\label{fig:hintcount}
\end{figure}

While for \sm{10.55} (\sperc{99.9}) domains IPv4 hints are available, \sm{10.23} (\sperc{96.9}) additionally advertise IPv6 addresses.
Most hints contain two addresses respectively but up to eight different addresses are visible as shown in \Cref{fig:hintcount}.
This allows a client to select from a set of different addresses and fallback to alternatives if necessary.
All other keys are only visible with a few domains.
The advertised ports in \httpsrr records are 80 (2$\times$), 443 (10$\times$) and 8920 (1$\times$).
Furthermore, we only receive 20 \ac{ech} configurations.
This supports the discussion that the respective \ac{ech} draft \cite{draft-ietf-tls-esni-16} still lacks deployment while the \acp{dnsrr} are already deployed for many domains and both drafts should be decoupled. \cite{logjam}
\sk{146.5} domains from the  Alexa~\cite{alexa}, \sk{169} from Majestic~\cite{majestic} and \sk{80.8} domains from the Umbrella~\cite{umbrella} \topm lists have an \httpsrr record.
The most prominent candidates are \texttt{google.com} with a service mode record and an \ac{alpn} parameter \textit{h2,h3} and \texttt{youtube.com} with a service mode record without additional data.

\noindent
\textit{\textbf{Key take-away:}
The \svcbrr record in both scenarios are only deployed by few domains.
In contrast, more than \sm{10} domains make use of \httpsrr records, mostly serving address hints and \ac{alpn} values.
The alias mode or remaining parameters are rarely used and should be reevaluated in the future.
}

\subsection{Involved Operators}
For the following analysis, we focus on domains with \httpsrr records in service mode (\sm{10.55}) due to their advanced deployment.
To get a better understanding of involved operators, we analyze where domains are hosted and which name servers are used.
If available, we use \textit{ipv4hints} and map addresses to the \ac{as} announcing the respective prefix.
For all domains without this parameter, we use queried A records for IPv4 addresses.

Domains with \httpsrr records are hosted in \sk{2.3} \acp{as}.
However, \Cref{tab:web_hosting} shows that a majority of domains (\sperc{98.8}) resolves to \acp{as} operated by Cloudflare (AS13335 and AS209242).
Domenshop, a Norwegian web hoster, hosts the second-highest number of domains and accounts for a large share of domains indicating support for \ac{http}/2 but not \ac{http}/3.
Following the Top 3 a more even distribution of the remaining \sk{72} domains across \sk{2.3} \acp{as} is visible.

To analyze responsible name servers, we rely on NS records for domains exactly matching domains in our input.
We do not follow CNAME records or extract information from SOA records.
During our scan, we received NS records for  \sm{7.8} domains with an \httpsrr record.
Domains without NS records in our data are either resulting in SOAs only (mostly \textit{www.} domains) or resolve to canonical names and would require further resolution steps.
In general, we are able to identify name servers supporting \httpsrr records hosted in 661 different \acp{as}.
This shows a widespread deployment of name servers that support the new record in general.

Similar to web hosting, most \httpsrr records are served by name servers hosted within Cloudflare followed by Domenshop.
The latter appears as three different \acp{as} (AS1921, AS12996, AS208045).
Each \ac{as} hosts a name server authoritative for a similar amount of domains respectively. 
Most domains have one NS record for each of the three name servers for resilience.

\noindent
\textit{\textbf{Key take-away:}
Domains with \httpsrr records are hosted in more than \sk{2.3} \acp{as} and name servers serving the records are in more than \sk{1.6} \acp{as}.
However, most records are hosted in and served by Cloudflare (\sperc{98}).
}

\begin{table}
	\centering
	\caption{Top 5 web hosters (out of \sk{2.3}) and name server providers (out of 661) of domains with \httpsrr records.}
	\label{tab:web_hosting}
	\begin{threeparttable}
		\setlength{\tabcolsep}{4.8pt}
		\footnotesize
		\begin{tabular}{llr|llr}
			\toprule
			\multicolumn{3}{c|}{Hosting}  & \multicolumn{3}{c}{Name server} \\
			ASN & Name &  \#Doms & ASN &                             Name &  \#Doms \\
			\midrule
			13335  & Cloudflare &  \sm{10.4} & 13335 &               Cloudflare &  \sm{7.7}\\
			12996  & Domenshop &    \sk{61.6} & 12996\tnote{1} &     Domenshop &    \sk{24.0} \\
			209242 & Cloudflare    & \sk{49.7} &16509 &                   Amazon &     \sk{3.2} \\
			397273 & Render     & \sk{4.9} &397226 &            Neustar &     \sk{3.1} \\
			14061  & Digitalocean & \sk{4.6} &44273 &                 GoDaddy &     \sk{2.5} \\
			\bottomrule
		\end{tabular}
		\begin{tablenotes}
			\item[1] Domenshop uses three different name servers for most domains located in three different \acp{as} (AS1921, AS12996, AS208045)
		\end{tablenotes}
	\end{threeparttable}
\end{table}

\subsection{Validity of Records}
We conducted \ac{http} scans to check the validity of collected records and whether clients can use the received information for an \ac{http} request.
The general scan approach is described in \Cref{sec:data}.
We focus on \ac{http}/1.1, \ac{http}/2 and \ac{http}/3, and select targets for each scan based on the \ac{alpn} and IP address hints.
\Cref{tab:scans} provides an overview about results.
\ac{tls}/\ac{tcp} handshakes are successful for more than \sperc{96.6} of evaluated targets for each \ac{http} version respectively while \quic handshakes are successful for more than \sperc{93.6} of \ac{http}/3 targets.
For \sperc{90}, we are further able to conduct an \ac{http} HEAD request.
Most unsuccessful connection attempts either result in a time out (1.1: \sk{6.4}, 2: \sk{9.1}, 3: \sk{50.8}) or a generic \ac{tls} handshake failure (1.1: \sk{708.9}, 2: \sk{692.1}. 3: \sm{1.2}).

Successful scans for \ac{http}/1.1 and \ac{http}/2 still cover \sk{1.8} \acp{as} while \ac{http}/3 and thus \quic scans only cover 416 \acp{as} out of \sk{2.3} candidates.
Analyzing failed scans reveals that a major origin of errors during \quic scans and for timeouts during the \ac{http} request is an attack prevention mechanism by Cloudflare \cite{cloudflare_under_attack}.
It is an automated challenge mechanism that delays the page load which results in errors with both the Goscanner and \qscanner.

Furthermore, we find \sk{23.0} domains with \httpsrr records served by Cloudflare name servers but hosted in different \acp{as} that only result in timeouts at least during \quic scans.
For those domains, scan results (timeouts) are reproducible.
Interestingly, those domains are hosted in more than \sk{1.3} \acp{as} and no relation is visible besides the Cloudflare name servers.
Furthermore, all \httpsrr contain the same \ac{alpn} set (\textit{h3, h3-29, h2}).
We assume a misconfiguration and informed Cloudflare.

\noindent
\textit{\textbf{Key take-away:}
A majority of available \httpsrr records contains valid, usable information especially if used by clients able to pass Cloudflare's attack prevention.
However, we identify a set of records with incorrect \ac{alpn} values. 
For those domains requests for some announced \ac{alpn} values time out consistently (mostly \ac{http}/3).
}

\begin{table}
	\centering
	\footnotesize
	\caption{Protocol scan results based on \httpsrr records. Targets are a combination of domain and IP address pairs.}
	\label{tab:scans}
	\begin{tabular}{lrrrrr}
		\toprule
		&& \multicolumn{4}{c}{Successful} \\
		\cmidrule(lr){3-6}
		\ac{http}  & Targets & \multicolumn{2}{c}{\ac{tls} Handshake} & \multicolumn{2}{c}{\ac{http} Requests} \\
		\midrule
		1.1 & \sm{21.44} & \sm{20.72} & \sperc{96.63} & \sm{19.48} & \sperc{90.84} \\
		2 & \sm{21.43} & \sm{20.73} & \sperc{96.69} & \sm{19.47} & \sperc{90.84} \\
		3 & \sm{19.59} & \sm{18.34} &  \sperc{93.64} & \sm{17.04} & \sperc{87.01} \\
		\bottomrule
	\end{tabular}
\end{table}

\section{Related Work}
\label{sec:related}
\svcbrr and \httpsrr records have seen little attention by other research so far.
In 2021, Zirngibl \etal~\cite{zirngibl2021over9000} used \httpsrr records to identify \quic deployments.
They found records for \sm{2.9} domains indicating \quic support hosted in \sk{1.2} \acp{as}.
However, they do not analyze records further.
In contrast, they find \ac{http} \ac{altsvc} Headers for more than \sm{20} domains.
While the latter is an alternative approach to distribute endpoint information, it requires a previous \ac{http} communication.
Two years later, we find 4$\times$ more \httpsrr records hosted in twice as many \acp{as}.
Similarly, Trevisan \etal \cite{trevisan2021altsvc} use alternative service information to identify \quic deployments but only \altsvc headers from additional \ac{http} requests.
Both, Zirngibl \etal and Trevisan \etal implied that \ac{http} \ac{altsvc} Headers are widely deployed.
We show that still fewer \httpsrr records are deployed, but growth is visible.

In 2019, Chai \etal~\cite{chai2019esni} evaluated Encrypted \ac{sni}, an older version of \ac{ech} that relied on TXT \ac{dnsrr} to distribute key information.
They identified more than \sk{100} domains within the Alexa \topm.
Similar results have been reported by Tsiatsikas \etal~\cite{Tsiatsikas2023measuringECH} in 2022.
In 2022, Hoang \etal~\cite{hoang2022esni} find \sperc{1.5} to \sperc{2.25} domains with a respective TXT record out of \sm{300} domains from TLD zone files. 
We show that no transition to \ac{ech} and \httpsrr records is visible yet.
Weber~\cite{oarc_akamai} reported about the visibility of \httpsrr queries from a network (Akamai) perspective.
While many queries failed with incorrect behavior initially, the correctness of seen responses changes quickly.
Additionally, they only observed records for \sk{126.4} domains and no alias mode.
Aguilar-Melchor \etal~\cite{aguilar2023turbotls} evaluate a potential positive effect of \httpsrr records but do not evaluate its current deployment state.

Furthermore, the security and impact of \ac{ech} has been analyzed~\cite{Bhargavan2022symbolicECH,Shamsimukhametov2022ECHclassification} and related work has evaluated the state of \ac{dns} over TCP, \ac{http} or \quic~\cite{kosek2022dnstcp,boettger2019doh,kosek2022doq,doan2021dot}, and shows increased deployment and in general good performance.
Thus, the fundamentals for a successful deployment of \svcbrr and \httpsrr records are given.
\section{Conclusion}
\label{sec:conclusion}
In this work, we provide the first large-scale overview of the deployment of new \svcbrr and \httpsrr \ac{dns} resource records.
While we find only very few domains with \svcbrr records (\sk{3.96} without and 26 with an Attrleaf label), we show that more than \sm{10} domains already resolve to \httpsrr records.
These records mainly provide \ac{alpn} values and \textit{ipv4-} and \textit{ipv6hints}.
We find only 20 domains with an \ac{ech} parameter which indicates lacking deployment.
However, we show that most domains are hosted within Cloudflare, and Cloudflare operated name servers are authoritative.

Nevertheless, information contained in most available records is correct, and handshakes followed by \ac{http} requests with indicated versions are possible.
Therefore, clients already querying the records (e.g., Apple devices~\cite{apple_announcement}) can effectively make use of \httpsrr records for more than \sm{10} domains and reduce, \ac{dns} requests and visible meta-data during connections establishments while reducing handshake cost.

\section*{Acknowledgment}%
\label{sec:acknowledgment}
The authors would like to thank the anonymous reviewers for their valuable feedback. This work was partially funded by the German Federal Ministry of Education and Research under the project PRIMEnet (16KIS1370), 6G-life (16KISK002) and 6G-ANNA (16KISK107) as well as the German Research Foundation (HyperNIC, grant no. CA595/13-1). Additionally, we received funding by the Bavarian Ministry of Economic Affairs, Regional Development and Energy as part of the project 6G Future Lab Bavaria and the European Union’s Horizon 2020 research and innovation program (grant agreement no. 101008468 and 101079774).

\bibliographystyle{plain}
\bibliography{references}

\appendices

\end{document}